\documentclass[sigconf]{acmart}

\copyrightyear{2022}
\acmYear{2022}
\setcopyright{rightsretained}
\acmConference[MSR '22]{19th International Conference on Mining
Software Repositories}{May 23--24, 2022}{Pittsburgh, PA, USA}
\acmBooktitle{19th International Conference on Mining Software
Repositories (MSR '22), May 23--24, 2022, Pittsburgh, PA, USA}
\acmDOI{10.1145/3524842.3528506}
\acmISBN{978-1-4503-9303-4/22/05}

\usepackage{seqsplit}
\begin{document}
\title{Exploring Apache Incubator Project Trajectories with APEX}

\author{Anirudh Ramchandran*}
\email{aniramch@ucdavis.edu}
\affiliation{%
  \institution{University of California, Davis}
  \streetaddress{1 Shields Ave}
  \city{Davis}
  \state{California}
  \country{USA}
  \postcode{95616}
}

\author{Likang Yin*}
\thanks{*The authors contributed equally to this work.}
\email{lkyin@ucdavis.edu}
\affiliation{
  \institution{University of California, Davis}
  \streetaddress{1 Shields Ave}
  \city{Davis}
  \state{California}
  \country{USA}
  \postcode{95616}
}

\author{Vladimir Filkov}
\email{vfilkov@ucdavis.edu}
\affiliation{
 \institution{University of California, Davis}
 \streetaddress{1 Shields Ave}
  \city{Davis}
  \state{California}
  \country{USA}
  \postcode{95616}
}

%%
%% By default, the full list of authors will be used in the page
%% headers. Often, this list is too long, and will overlap
%% other information printed in the page headers. This command allows
%% the author to define a more concise list
%% of authors' names for this purpose.

%%
%% The abstract is a short summary of the work to be presented in the
%% article.
\begin{abstract}
Open Source Software (OSS) is a major component of our digital infrastructure, yet more than 80\% of such projects fail.
Seeking less uncertainty, many OSS projects join established software communities, e.g., the Apache Software Foundation (ASF), with established rules and community support to guide projects toward sustainability.
In their nascent stage, ASF projects are incubated in the ASF incubator (ASFI), which provides systematic mentorship toward long-term sustainability.
Projects in ASFI eventually conclude their incubation by either graduating, if successful, or retiring, if not. 

Time-stamped traces of developer activities are publicly available from ASF, and can be used for monitoring project trajectories toward sustainability.
Here we present a web app dashboard tool, APEX, that allows  internal and external stakeholders to monitor and explore ASFI project sustainability trajectories, including social and technical networks.
%Through use cases we demonstrate APEX's utility in monitoring for project downturn events, identifying longer term engagements, and project comparison. The APEX app is available at \url{https://anirudhsuresh.github.io/APEX}. The code and data are available at \url{https://github.com/anirudhsuresh/APEX/}.
\end{abstract}

\begin{CCSXML}
<ccs2012>
<concept>
<concept_id>10011007.10011074.10011134.10003559</concept_id>
<concept_desc>Software and its engineering~Open source model</concept_desc>
<concept_significance>500</concept_significance>
</concept>
</ccs2012>
\end{CCSXML}

\ccsdesc[500]{Software and its engineering~Open source model}

\keywords{OSS Sustainability; Apache Incubator; Tool}

%\ccsdesc[500]{Software and its engineering~Open source model}

%%
%% Keywords. The author(s) should pick words that accurately describe
%% the work being presented. Separate the keywords with commas.
%\keywords{Dashborad Tool, Apache Software Foundation, OSS Sustainability}

%%
%% This command processes the author and affiliation and title
%% information and builds the first part of the formatted document.
\maketitle

\section{Introduction}
%Open Source Software (OSS) is ubiquitous in our modern digital society, and is supported by individual, corporate, and government efforts. 
%The daily operations of countless individuals, big companies, and national governments rely on OSS.
In spite of the large amounts of resource put in them, many OSS projects end up on trajectories that are ultimately not sustainable. 
%Then, from a societal perspective, it is important to ask: Can we help the OSS projects that could have been sustainable? And how?
In recent work we showed that OSS project sustainability can be effectively predicted early on in project development from longitudinal project and process metrics supplemented by socio-technical network metrics (developer communications and code contributions), specific to the Apache Software Foundation (ASF)
~\cite{yin2021forecasting}. 
ASF, as one of the most popular OSS communities, provides specific guidelines and establishes regulations to help OSS projects eventually become self-sustainable.
Nascent projects with ASF aspirations are housed in the Apache Software Foundation Incubator (ASFI) for a period of time, after which they are graduated into ASF if they are found to be sustainable, otherwise they get retired.

Promisingly, our work~\cite{yin2021forecasting} implied that monitoring and reflecting on their sustainability forecast can enable projects to act proactively, and potentially  correct downturns in the forecasts.
To enable such monitoring in practice, here we present a dashboard tool, {\em{APEX}}, intended for nascent projects in the Apache Software Foundation Incubator to monitor their sustainability trajectories over time, thus allowing for timely course corrections and for potentially  improving the likelihood of project graduation into ASF.
Our motivation goes beyond ASFI as many nascent OSS projects fall outside the ASF domain, and its well developed community support structure. Self-monitoring and self-correction may be even more pertinent to those.
While intended for ASF projects specifically, APEX is designed in a generic way and thus can easily accommodate data from repositories other than ASF.

\noindent\underline{Related Work}
ASF provides a monitoring tool, Clutch\footnote{Clutch Analysis: \url{http://incubator.apache.org/clutch/}}, to help OSS developers self-reflect and take actions when their projects are experiencing issues. 
The Clutch tool uses colors to signal the status of project metrics, e.g., missing documentation, lack of new committers, etc.
Although it works well for its intended use, Clutch's analysis is of limited use as a real-time monitoring tool since
(1) it is not project-specific, 
i.e., all projects follow the same standards for all features regardless of their project size or context;
(2) it does not consider historical records; and 
%OSS projects development often shows trends, monitoring projects longitudinally is of importance.
(3) it does not make actionable suggestions.
%The Clutch analysis does not provide structural overview of the project, thus hard to take actions when downturns come, e.g., who are collaborating on what tasks?
Our APEX tool complements the existing Clutch tool by providing additional analytics power for understanding the longitudinal socio-technical aspects of projects. It also can yield potential actionable insights. 

% \textbf{OTHER TOOLS (PROVIDE 3-4 PAPERS)}
There are other projects that focus on analytics for OSS sustainability outside of ASF domain. E.g., the Augur and GrimoireLab within the CHAOSS (Community Health Analytics Open Source Software) project\footnote{CHAOSS community \url{https://chaoss.community/}}, provide a toolbox for project sustainability self-monitoring. 
However, unlike APEX, they don't provide a synthesis of metrics into a longitudinal sustainability forecasts, or allow deep dives into email and commits.
%OSS developers on GitLab are also self-organized themselves to help OSS projects become long-term sustainable by building tools recognizing all types of contributions instead of only coding, e.g., documentation and volunteer management\footnote{\url{https://gitlab.com/gitlab-org/gitlab/-/issues/327138}}.

Next, we first introduce APEX, and then describe use cases that demonstrate its utility in a) monitoring for ASF project downturn events, b) identifying longer term engagements between developers, and c) within-ASF project comparisons. The APEX app is available at \url{https://ossustain.github.io/APEX/}. Code and data are available at \url{https://github.com/ossustain/APEX/}.
\begin{figure}[t]
    \centering
    \includegraphics[width= 0.9\linewidth ]{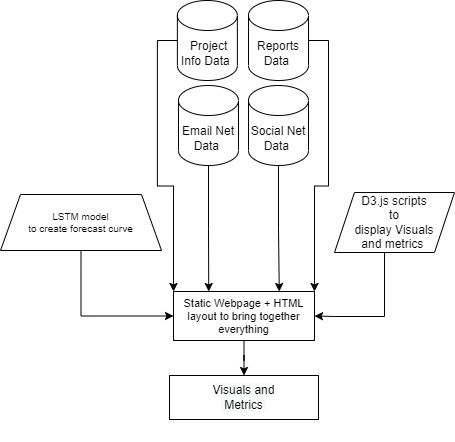}
    \caption{The APEX pipeline}
    \label{fig:1}
    \vspace{-0.2in}
\end{figure}

\section{Data and Implementation}

\begin{figure*}[t]
    \centering
    \includegraphics[width=0.70\linewidth]{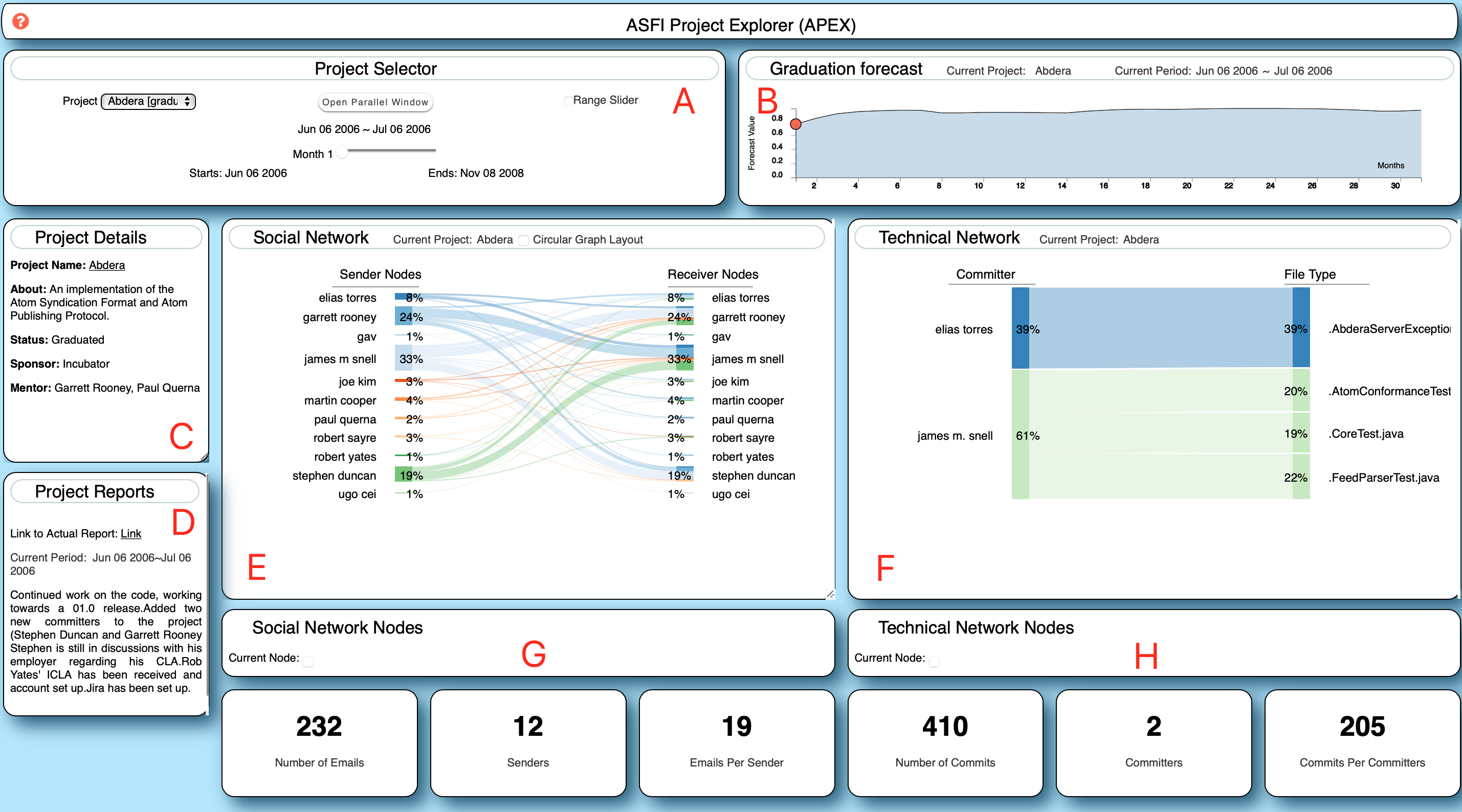}
    \caption{Layout of the APEX Dashboard}
    \label{fig:2}
\end{figure*}

\underline{Data Source}
The APEX pipeline depends on four types of data: basic project information, periodic project reports, email communications, and code commits.
We use our previously published dataset from ASFI~\cite{yin2021apache} to obtain the emails and commits.
The dataset comprised 211 graduated and 62 retired incubator projects (in total of 273), with 1,201,746 emails, and 3,654,196 commits, with each project spending on average 22.32 months in incubation.

We scraped project information (name, mentors, dates, status) from the project's ASF Incubator homepage\footnote{ASF incubator: \url{http://incubator.apache.org/projects/}}, using Python’s BeautifulSoup Package.
The Apache mailing list archive,%\footnote{ASF mailing list archive: \url{http://mail-archives.apache.org/mod_mbox/}},
contains full historical information for all projects including project participants, mentors who assisted with the projects, project reports, all emails, and all commits.
%Our ASFI data is organized starting from the month it began the incubator to end of their incubation or its most recent month if they are still in incubation. 
To obtain developer emails we frequently had to backfill partial email addresses by writing scripts to search for the partial email throughout the  email text.
We used the same approach to identify unique committers that may have used aliases.

\underline{Social and Technical Networks}
From the ASF incubator data we derive two kinds of longitudinal networks for each project, for each month: a social and a technical. The social networks have directed edges between developer, derived from the email archives, using the method by Bird et al.~\cite{bird2006mining}. 
We present the social networks in a bipartite graph layout, using Sankey diagram~\cite{riehmann2005interactive}.
%two types of layouts: a force-directed graph and a bipartite network.
%In the former, the active developer names are displayed as nodes of the graph, for a given month. 
There are two sets of nodes: on the left are all senders of messages in a given month. On the right are those who either received a direct message, or, in the case of a broadcast message, those who have responded to that message.
The edges are directed from the sender to the receiver node, except for the broadcast messages where the edges are directed from a node that has sent an email to a node that has replied to that email~\cite{bird2006mining}. 

The technical network is also a bipartite graph with two sets of nodes: on the left are the developers that have made commits in a given month and on the right are all the file types committed to in that month (e.g., .java, .html, etc., based on their file extensions).
The edges connect the developers to the file types they committed to in that month. 
%These networks are visualized using D3.js. 
We aggregate the network edges in a monthly manner, i.e., the longitudinal networks of each project consist of monthly network snapshots.
%The descriptive statistics for the networks are also given below the networks.

\underline{Implementation Technology}
The APEX tool design and pipeline is illustrated in Figure 1. 
To implement the front-end of APEX we use the {\em D3} JavaScript library (\url{https://d3js.org/}).  \textit{D3} uses visual layouts and an associated tool-set to improve front-end efficiency. 
\textit{D3} also provides developers with design flexibility through standardized data manipulation operations. 
Additionally, one of D3's foci is transitions and animation. 
This allows our dashboard to quickly adapt to changing inputs, e.g., a change of the current month.
%Many JavaScript third-party libraries are implemented based on two rendering technologies: Canvas and SVG. 
%Canvas is a pixel-based library, and it generally does not respond after drawing figures, i.e., it does not prefer user interactions. 
To provide interactivity of our tool, we use the SVG rendering technology, which is based 
on Document Object Model (DOM) operations and supports precise user interaction. 
We also make use of {\em jQuery} extensively. 
jQuery is a JavaScript library that helps simplify and standardize interactions between JavaScript code and HTML elements. 
%JavaScript allows websites to be interactive and dynamic, and jQuery streamlines that process. 
We use jQuery to design event listeners: processes in JavaScript that wait for an event to occur. 
%A simple example of an event is a user clicking the mouse in a particular section or pressing a key on the keyboard.
This has allowed us to create a seamless dynamic experience throughout the dashboard, where a change in one section adjusts all related sections and visuals accordingly.
We have also made use of a lightweight range slider with multi-touch support called {\em noUIslider}, and it has an in-built event listener function allowing integration with the rest of the elements in the dashboard. 
Frequent DOM operations are costly, negatively impacting the user experience by screen flashing and stuttering during the interactions. 
We relax this cost by keeping each project and month stored in separate json files. 
%Our dashboard-based tool APEX is available at \url{https://anirudhsuresh.github.io/APEX/}.

\underline{Sustainability Forecasting} APEX features the AI based sustainability forecasting model by Yin et al.~\cite{yin2021forecasting}. 
They implemented a 3-layer LSTM model: a $64$ neurons LSTM layer with a $0.3$ rate drop-out layer, and then followed by a dense layer with the \textit{softmax} function to yield the predicted likelihood of project graduation. 
In the experimental setup, the graduated projects are encoded as 1; retired projects as 0. 
During training, the monthly socio-technical networks variables (e.g., number of nodes/edges, clustering coefficient, and mean degree in the networks) of each project were fed into the model.
This LSTM neural network based model gives a sustainability forecast in each month of the project development. 
More experimental details can be found in the paper~\cite{yin2021forecasting}.

%The sustainability forecast, combined with socio-technical networks, can be used for dynamic monitoring OSS projects' progress to build a sustainable community. 

%\underline{Using the data effectively.}
%Frequent DOM operations are costly, negatively impacting the user experience by screen flashing and stuttering during the interactions. 
%To avoid this, we keep each project and month stored in a separate file. 
%If the page needs to render excessive DOM elements simultaneously, it only displays the most significant ones, which is the selected <project, month> pair of interests. We load and use only the project and month that the user selects. All previous DOM elements are cleared when the user changes the project or month.Additionally in the case of multiple months being selected, we wait for the user to select the month range and only then render the necessary changes to display them. The above techniques improve the overall-user experience

\section{Dashboard Elements}

\underline{Dashboard Panes.}
There are four main sections to the dashboard, see Figure 2:  (1) top pane (panes A,B), (2) left pane (panes C,D), (3) middle pane (panes E and G) (4) right pane (panes F and H).

The {\em top pane} has two parts to it. The left side (A) allows the user to select a project of interest in the drop-down menu and a specific month through the month slider. 
Based on these inputs, the other panes change dynamically to display the respective information for the selected project and month. 
In addition, this section also allows the user to toggle a checkbox to switch to a range slider to display a range of months, e.g., 1-5 months, instead of a single month. 
To the right (B) is the sustainability forecast visual, which depicts the sustainability forecast for the project, ranging from 0 (not sustainable) to 1 (sustainable), for any given month. 

Below those, the left pane consists of two distinct sub-sections: the project info pane (C) and the project report pane (D). The former shows the project name, a link to the official website, and the project's status (i.e., graduated or retired). The ASF sponsor's name (if anyone in particular, otherwise `incubator') is displayed below that. At the bottom is a short introduction to the selected project.
Below that, in the project's report pane (D), we present the report submitted by the project to the ASFI, for the given month.

The {\em middle pane} consists of the social network visual on top (E) and its related metrics below it (G).
%As mentioned previously, the social network can toggle between a  and a force-directed graph. 
%The force-directed graph shows the network of email communications between developers in the selected month. The developers' names are the nodes in the network, placed in a circle, and the email communications form the edges. Developers who have also committed in that month are marked with a star. The edges have a direction (i.e., who sent the email and who received it), where the direction is indicated by color when the cursor hovers over a name: red stands for sending emails, and blue for receiving emails.
The social bipartite graph is presented as a Sankey diagram~\cite{riehmann2005interactive}, the height of a node illustrating the \% (relative to the total) of emails sent (left) or received/replied-to (right) by that developer in a given month. 
%On the left is a list of developers who sent a broadcast, while on the right is a list of developers who responded to the said broadcast. 
The sizes of the flows sre proportional to the number of emails exchanged between the developers. 
Hovering over a developer's name emphasizes all developers that have received a directed email, or responded to a broadcast, from that developer.

In the {\em right pane}, on top is the technical network of developers who have committed to files (F), and their metrics are below (H). For the visual, we also use a bipartite Sankey diagram. On the left is a list of developers, while the file type (i.e., extension) of the files committed to is on the right. 
The percentage of one's efforts relative to others' is shown on the left, with the sizes of the flows proportional to that.
%Similarly, for attention, the file types get on the right.
Hovering over a developer's name, or, respectively, a file type, emphasizes additional information: all file types a developer has touched, or all developers that have touched that file type along with their \% contribution, respectively.

Additionally, by hovering over and clicking on a developer's name in the social, respectively the technical networks, we get a button with their name under the network, which when clicked opens a window with a list of all their emails, respectively commits, in that month. They appear in a pop-up window next to the dashboard.

\section{Use Case Examples}
APEX conveniently shows in one place the project info, monthly aggregated code commits, email communications, and two unique features of ASF, the periodic report info and the graduation status. 
Such rich and fine-grained information can enable researchers and practitioners to study the trajectory of a given project by showing changes over time, including identifiable patterns and up/down trends, in the socio-technical networks and the sustainabiity forecasts.

\begin{figure}[tbp]
    \centering
    \includegraphics[width= \linewidth]{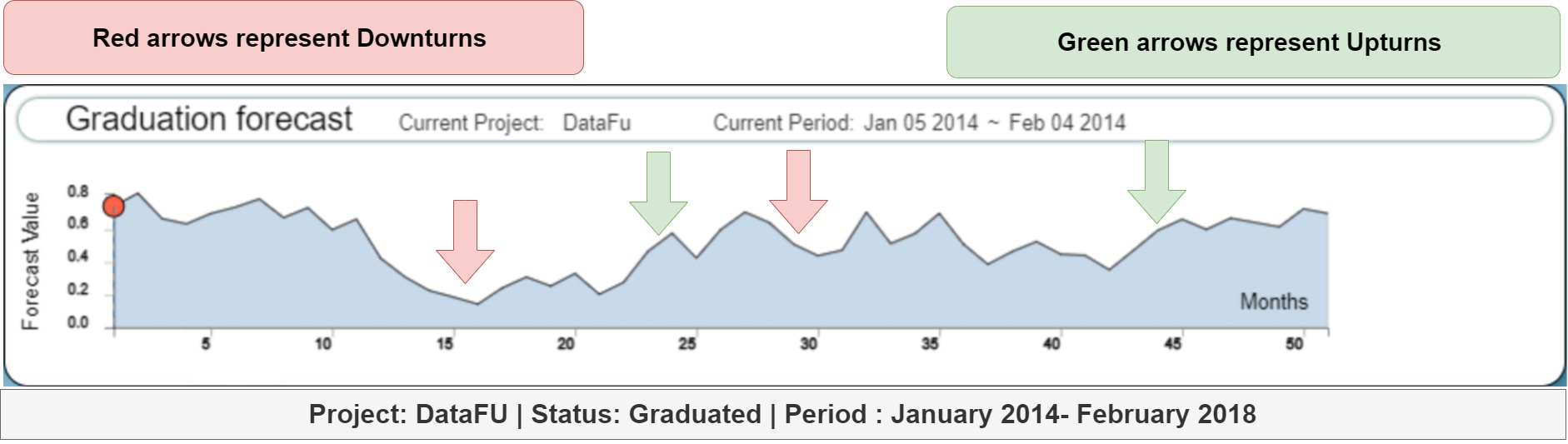}
    \caption{Using the sustainability forecast to understand and explore downturns (red) and upturns (green) of project \textit{DataFu}.}
    \label{fig:turns}
\end{figure}

\noindent\underline{\textbf{Use Case I: Studying Sustainability Turning Points}}
Patterns and trends in the longitudinal socio-technical networks can be studied to identify causes for downturns in the sustainability forecast, allowing APEX users to be proactive with changing project trajectories. 
For example, as shown in Figure~\ref{fig:turns}, for the selected project \textit{DataFu}, by simply eyeballing we can identify that there is a big downturn around month 12. 
Some possible reasons for this could be that
(1) The project just launched a big release; or
(2) Some core developers left the projects.
Going through the email discussions in months adjacent to the downturns may offer reasons for the changes, which in this case is likely the latter. 
%Similarly, we can also specify the underlying reasons for the upturns whenever it occurs.

Thus, the monthly social-technical networks combined with the real-time sustainability forecast can allow practitioners and researchers to monitor for downturn events and react proactively.

\underline{\textbf{Use Case II: Studying Different Length Engagements}} APEX allows aggregating the networks over a range of months. 
This allows the study and comparison of different length engagements, both social and technical in nature.
This can be done by enabling the range slider, which allows multiple months to be selected at once, yielding a time range for the nodes and edges in the networks. 
Once the range is specified, the metrics and the visuals are adjusted to display multiple consecutive months of interactions.

\begin{figure}[t]
    % \centering
    % \includegraphics[scale=\linewidth]{case2.jpg}
    \includegraphics[width= 0.9\linewidth]{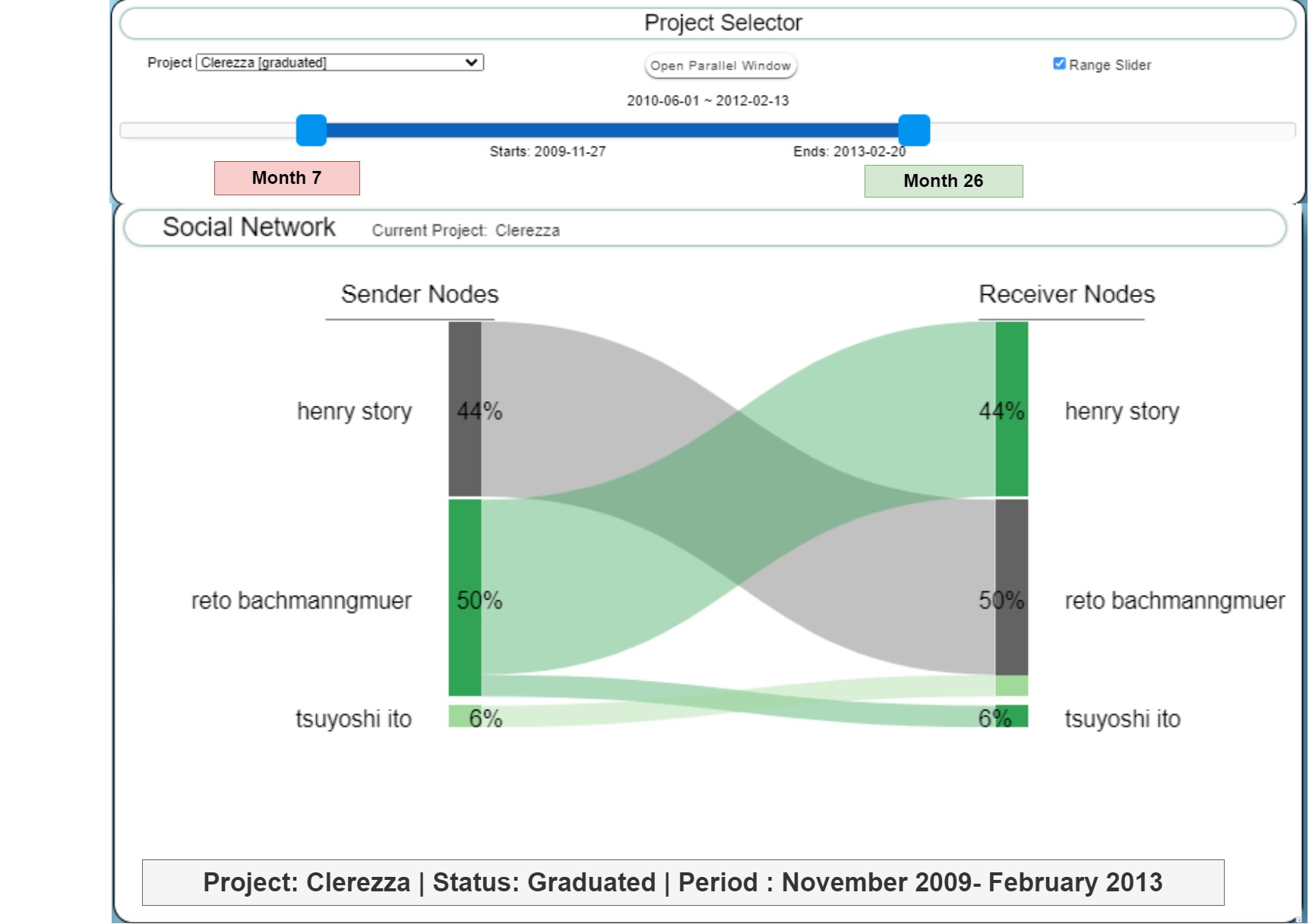}
    \caption{Aggregating project Clerezza's social networks over a range of months (7-26) shows longer-term engagements}
    \label{fig:longterm}
\end{figure}

An example of a social network over a longer range is shown in Figure~\ref{fig:longterm}.
There we see thicker and thinner edges; the former indicate communications that recur over multiple months, attesting to a longer term engagement between those developers, i.e., recurring communication.
%Short-term networks change over time fluctuating more and are smaller than the longer-term aggregated networks. 
By comparing the short-term and long-term social-technical networks, we can identify  recurring patterns over longer periods of time during the project incubation.

\begin{figure}[t]
    \centering
    \includegraphics[width= \linewidth]{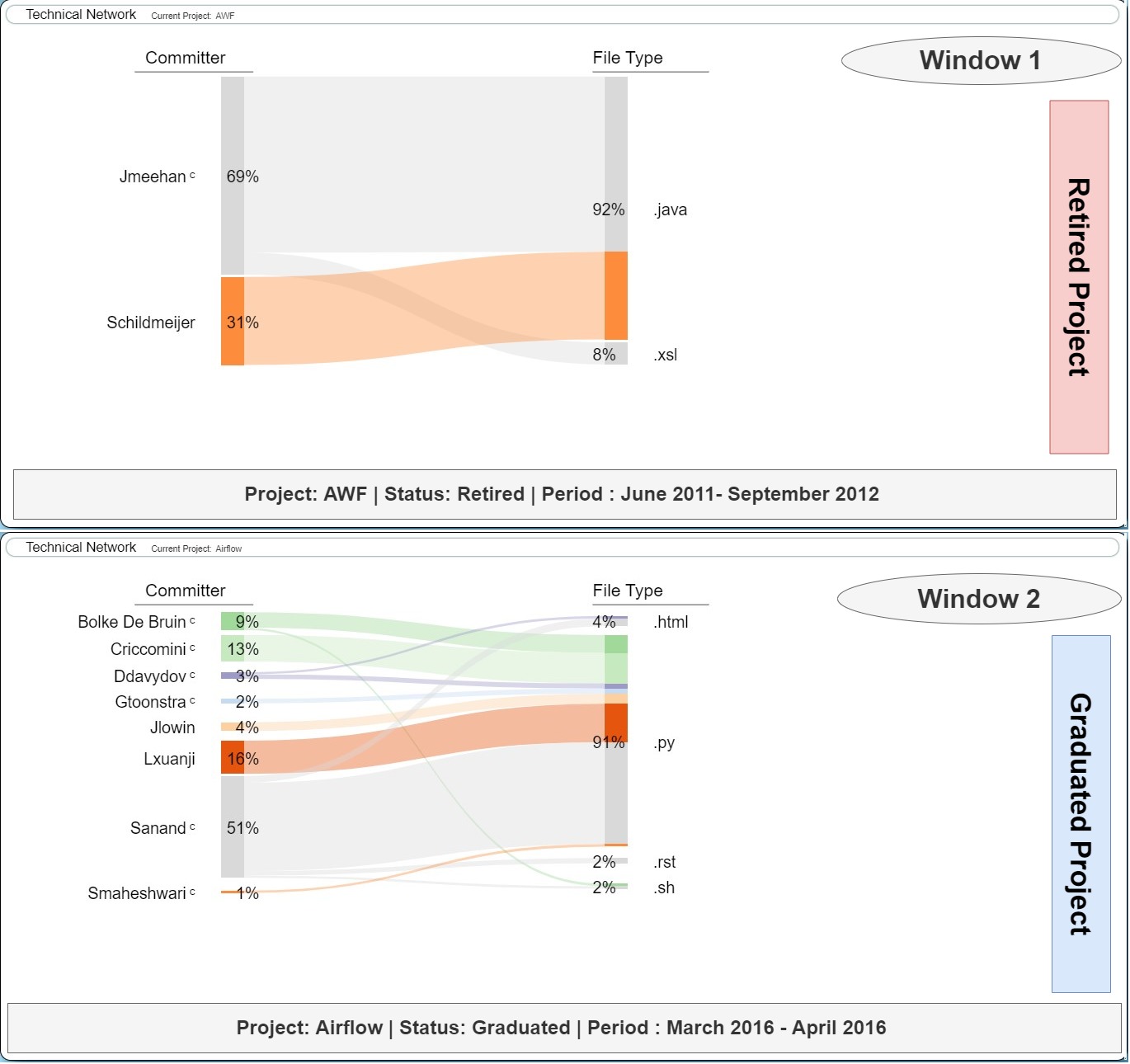}
    \caption{Parallel windows can contrast technical networks between project AWF (retired, top) and project Airflow (graduated, bottom)}
    \label{fig:5}
\end{figure}

\underline{\textbf{Use Case III: Cross-Project Comparison}}
APEX also allows users to compare and contrast two projects by opening up parallel windows of our dashboard.
%We could open up two or more projects in separate windows to explore relationships among multiple projects. 
Thereby, researchers can explore multiple projects simultaneously to generate hypotheses about relationships between their socio-technical structure and graduation status. 
%That is, we can pretty much tell whether a project will graduated or retired by only looking at first a few months data of the project. 
%APEX provides, even not in a statistical way, an approachable validation by taking out several comparable projects (e.g., they have similar size), and looking at how their network evolve over time.
E.g., Figure~\ref{fig:5} shows the technical networks of two projects, one graduated and the other retired. 
%An immediate pattern that emerges is that the communication in the retired project is more focused than in the graduated one.
An immediate pattern that emerges is that more developers are committing code changes in the graduated project than in the retired one.
%Then we can use APEX to compare the differences in their socio-technical networks month-by-month. 
Researchers can followup on this hypothesis by looking into the driving factors behind it, using, e.g., productivity studies, or topics of discussions. 
%APEX can present more detailed information by directing the users to the original discussions with email subject and email content. 
%Another use case could be to open up the same project but open up different months in the two windows to explore, for example, upturns and downturns in a particular project

\section{Using APEX Beyond ASF}
We have designed APEX to serve ASFI projects that are early in their incubation to monitor and reflect on their progress in a more agile way than previously possible. 
But APEX is in principle not limited to ASFI data. 
It takes as input JSON files and visualizes them in different ways.
To aid projects outside of ASF that want to benefit from it, we have made our full code and data publicly available. 
We provide a README file, \url{https://github.com/anirudhsuresh/APEX/blob/main/README.md}, that details the JSON formats of the required input data.
The README file links to scripts with which comma separated values (CSV) files, common outcome of repository mining, can be converted to the required JSON format for all the required APEX components: email networks, email metrics, technical networks, commit metrics, project info, project reports, and sustainability forecasts.
The last one will have to be calculated from the others, using the code provided in our previous study~\cite{yin2021forecasting}.

\section{Limitations and Conclusion}
\underline{Limitations} 
Our dataset is large and diverse (within ASF) but limited to ASFI projects, so generalizing beyond ASF is risky.
However, we provide a README file with instructions to aid non-ASF projects in using APEX.
%Additionally since all our current projects come from the Apache incubator, we acknowledge that some results can carry risks and we may not be able to generalize these results across different OSS projects.
Selecting a range of months can result in very dense networks that are hard to read or interpret. This function is most useful when limited to a few consecutive months.
 
%To mitigate this risk in the future we are looking to expand to more OSS projects from GitHub and other places.
%Also in order to add in new data to the dashboard our current data pipeline is long and disconnected and not automated.

\underline{Conclusion} 
Research into OSS project sustainability can present actionable insights for project maintenance. 
%However, there is a dearth of data that are dynamic and have extrinsic sustainability labels upon the exit of project incubation and tools that provide an ability to dig deeper into the reasons for sustainability. 
In this work, we presented a dashboard tool for exploring a longitudinal data-set of technical contributions and developer communication in ASF incubator projects, with extrinsic, graduation success labels and forecasts.
The tool can be used for real-time monitoring and study of ASFI projects.
It can also help generate hypothesis about OSS project sustainability.
Future work will be aimed at adding additional data sets.
We have engaged the CHAOSS project and are devising ways to standardize APEX and integrate it into CHAOSS.

\section*{Acknowledgement} This material is based upon work supported by the National Science Foundation under Grant No. 2020751.

\bibliographystyle{acm}
\bibliography{ref}
\end{document}